\documentclass[twocolumn]{revtex4}
\usepackage{dcolumn}
\usepackage{longtable}

\begin{document}

\title{Bohr Hamiltonian with Davidson potential for triaxial nuclei}

\author{I. Yigitoglu$^1$ and Dennis Bonatsos$^2$}

\affiliation{$^1$ Faculty of Science and Arts, Department of Physics, Gaziosmanpasa University, 60240 Tokat, Turkey} 
\affiliation{$^2$ Institute of Nuclear Physics, N.C.S.R. ``Demokritos'', GR-15310 Aghia Paraskevi, Attiki, Greece}

\begin{abstract}
A solution of the Bohr Hamiltonian appropriate for triaxial shapes, involving a Davidson potential
in $\beta$ and a steep harmonic oscillator in $\gamma$, centered around $\gamma=\pi/6$, is developed.
Analytical expressions for spectra and B(E2) transition rates ranging from a triaxial vibrator to
the rigid triaxial rotator are obtained and compared to experiment.
Using a variational procedure it is pointed out that the Z(5) solution, in which an infinite square well potential
in $\beta$ is used, corresponds to the critical point of the shape phase transition from a triaxial vibrator
to the rigid triaxial rotator.
\end{abstract}


\maketitle

\section{Introduction} 

The advent of critical point symmetries \cite{IacE5,IacX5,PPNPC,PPNPJ}, manifested experimentally \cite{CZE5,CZX5,CMcC} in nuclei
on or near the point of shape phase transitions, revived interest in special solutions \cite{Fort,Nature} of the Bohr
Hamiltonian \cite{Bohr}. While shape phase transitions in nuclei have been originally found \cite{Feng,PPNPJ} in the framework
of the Interacting Boson Model \cite{IA}, the first examples of critical point symmetries, the E(5) symmetry \cite{IacE5}
[corresponding to the second order phase transition between spherical and $\gamma$-unstable (soft with respect
to axial asymmetry) nuclei] and the X(5) symmetry \cite{IacX5} (appropriate for the first order phase transition between spherical
and prolate deformed nuclei), have been developed as special solutions of the Bohr Hamiltonian, using an infinite
square well potential in the $\beta$ degree of freedom (related to the magnitude of the deformation).
In E(5) the potential is independent of the $\gamma$ degree of freedom , related to the shape of the nucleus,
while in X(5) the potential is separable into two terms, $u(\beta)+v(\gamma)$, the latter being a steep harmonic
oscillator centered around $\gamma=0$, corresponding to prolate deformed nuclei. The Z(5) solution \cite{Z5} developed later,
formally resembles the X(5) case in using a separable potential and an infinite square well potential
in $\beta$,but it differs drastically in using a steep harmonic oscillator
potential in the $\gamma$ degree of freedom  centered around
$\gamma=\pi/6$, corresponding to triaxial shapes.

Triaxial shapes in nuclei have been considered for a long time, since the introduction of the rigid triaxial rotor \cite{DavFil,DavRos},
despite the fact that very few candidates have been found experimentally \cite{stag,stag2}. In the framework of the IBM, triaxial
shapes can occur in three different cases:

\noindent i) In the IBM-1 framework, in which no distinction
between protons  and neutrons is made, the inclusion of higher order (three-body) terms is needed \cite{Heyde,Thiamova}.

\noindent ii) In the IBM-2 framework, in which protons and neutrons are used as distinct entities,
the inclusion of one-body and two-body terms suffices \cite{AriasPRL,CIPRL,CIAP}.

\noindent iii) In the sdg-IBM framework, the presence of the g-boson also suffices \cite{PVIBZ}.

\noindent
Shape phase transitions involving rigid triaxial shapes have been studied recently in the IBM-2 framework \cite{AriasPRL,CIPRL,CIAP},
while in the sdg-IBM framework no transitions towards stable triaxial shapes have been found so far \cite{PVIBZ}.

In the present work, the Z(5) solution is modified by replacing the
infinite square well potential in $\beta$ by a Davidson potential
\cite{Dav},
\begin{equation}\label{eq:e0}
u(\beta)= \beta^2 + {\beta_0^4\over \beta^2},
\end{equation}
where $\beta_0$ corresponds to the position of the minimum of the
potential. This solution is going to be called Z(5)-D. Similar
studies already exist in the literature for both the E(5)
\cite{Elliott,Fort} and X(5) \cite{varPLB,varPRC} cases. In
addition, other potentials, like $\beta^{2n}$ potentials
\cite{Arias,BonE5,BonX5}, and the Morse \cite{Inci} and Kratzer
\cite{Fort} potentials have been used in the E(5)
($\gamma$-unstable) \cite{FV1} and X(5) ($\gamma\approx 0$, prolate
deformed) \cite{FV2} frameworks.

In addition to providing easily comparable to experiment analytical solutions for the spectra and B(E2) transition rates,
the present study  leads to an important by-product. Using a variational procedure applied earlier in the E(5)
and X(5) frameworks \cite{varPLB,varPRC}, one can see that the Z(5) solution can be interpreted as corresponding
to the critical point of a shape phase transition between a triaxial vibrator and a rigid triaxial rotator.

In Sections 2 and 3 the $\beta$ part and the $\gamma$ part of the spectrum are considered, while B(E2) transition
rates are calculated in Section 4. Numerical results, including results of the above-mentioned variational procedure,
are shown in Section 5, while in Section 6 a brief comparison to experiment is attempted. Finally, conclusions and plans
for further work are found in Section 7.

\section{The $\beta$-part of the spectrum} 

The original Bohr Hamiltonian \cite{Bohr} is
$$
H = -{\hbar^2 \over 2B} \left[ {1\over \beta^4} {\partial \over \partial
\beta} \beta^4 {\partial \over \partial \beta} + {1\over \beta^2 \sin
3\gamma} {\partial \over \partial \gamma} \sin 3 \gamma {\partial \over
\partial \gamma} \right.$$
\begin{equation}\label{eq:e1}
\left.  - {1\over 4 \beta^2} \sum_{k=1,2,3} {Q_k^2 \over \sin^2
\left(\gamma - {2\over 3} \pi k\right) } \right] +V(\beta,\gamma),
\end{equation}
where $\beta$ and $\gamma$ are the usual collective coordinates, while
$Q_k$ ($k=1$, 2, 3) are the components of angular momentum in the intrinsic frame,
and $B$ is the mass parameter.

In the case in which the potential has a minimum around $\gamma =\pi/6$,
the term involving the components of the angular momentum can be written \cite{Z5}
in the form $ 4(Q_1^2+Q_2^2+Q_3^2)-3Q_1^2$.
Using this result in the Schr\"odinger equation corresponding to
the Hamiltonian of Eq. (\ref{eq:e1}), introducing \cite{IacX5} reduced energies
 $\epsilon = 2B E /\hbar^2$ and reduced potentials $u = 2B V /\hbar^2$,
and assuming \cite{IacX5} that the reduced potential can be separated into two
terms, one depending on $\beta$ and the other depending on $\gamma$, i.e.
$u(\beta, \gamma) = u(\beta) + v(\gamma)$, the Schr\"odinger equation can
be approximately separated into two equations
$$ 
\left[ -{1\over \beta^4} {\partial \over \partial \beta} \beta^4
{\partial \over \partial \beta} + {1\over 4 \beta^2} (4L(L+1)-3\alpha^2)
+u(\beta) \right] \xi_{L,\alpha}(\beta)  $$
\begin{equation} \label{eq:e3}
 =\epsilon_\beta  \xi_{L,\alpha}(\beta),
\end{equation}
\begin{equation}\label{eq:e4}
\left[ -{1\over \langle \beta^2\rangle \sin 3\gamma} {\partial \over
\partial \gamma}\sin 3\gamma {\partial \over \partial \gamma}
+v(\gamma)\right] \eta(\gamma) =
\epsilon_\gamma \eta(\gamma),
\end{equation}
where $L$ is the angular momentum quantum number, $\alpha$ is the projection
of the angular momentum on the body-fixed $\hat x'$-axis
($\alpha$ has to be an even integer \cite{MtVNPA}),
$\langle \beta^2 \rangle$ is the average of $\beta^2$ over $\xi(\beta)$,
and $\epsilon= \epsilon_\beta +\epsilon_\gamma$. It should be noticed that the separation 
of variables is approximate, since in Eq. (\ref{eq:e4}) the quantity $\langle \beta^2\rangle$
appears, which depends on the quantum numbers $L$ and $\alpha$, appearing in Eq. (\ref{eq:e3}). 
Therefore an approximate separation of variables is achieved in the adiabatic limit, as in Ref. \cite{Bijker}.
As a consequence, the energy relation $\epsilon= \epsilon_\beta +\epsilon_\gamma$ is also approximate.

The total wave function should have the form
$ \Psi(\beta,\gamma,\theta_i) = \xi_{L,\alpha}(\beta) \eta(\gamma)
{\cal D}^L _{M,\alpha}(\theta_i)$,
where $\theta_i$ ($i=1$, 2, 3) are the Euler angles, ${\cal D}(\theta_i)$
denote Wigner functions of them, $L$ are the eigenvalues of angular
momentum, while $M$ and $\alpha$ are the eigenvalues of the projections
of angular momentum on the laboratory fixed $\hat z$-axis and the body-fixed
$\hat x'$-axis respectively.

Instead of the projection $\alpha$ of the angular momentum on the
$\hat x'$-axis, it is customary to introduce the wobbling quantum number
\cite{MtVNPA,BM} $n_w=L-\alpha$, which labels a series of bands
with  $L=n_w,n_w+2,n_w+4, \dots$ (with $n_w > 0$) next to the ground state
band (with $n_w=0$) \cite{MtVNPA}.

Eq. (\ref{eq:e3}) has been solved
in the case in which $u(\beta)$ is an infinite well potential,
the corresponding solution called Z(5) \cite{Z5}. The spectrum is given
by roots of Bessel functions, for which the notation has been kept the same as in Ref.
\cite{IacX5}, namely $E_{s,n_w,L}$, the ground state band corresponding to $s=1$, $n_w=0$.

Eq. (\ref{eq:e3}) is exactly soluble also in the case in which
the potential has the form of a Davidson potential \cite{Dav}
\begin{equation}\label{eq:e12}
u(\beta)= \beta^2 +{\beta_0^4\over \beta^2},
\end{equation}
where $\beta_0$ is the position of the minimum of the potential.
When plugging the Davidson potential in Eq. (\ref{eq:e3}), the
$\beta_0^4/\beta^2$ term of the potential is combined with the
$[4L(L+1)-3\alpha^2]/4\beta^2$ term appearing there and the
equation is solved exactly \cite{Elliott,Rowe}, the eigenfunctions being
Laguerre polynomials of the form
$$
\xi_{n,n_w,L}(\beta) =\xi_{n,\alpha,L}(\beta)=$$
\begin{equation}\label{eq:e13}
\left[ {2 n!\over \Gamma
\left(n+a+{5\over 2}\right)}\right]^{1/2} \beta^a L_n^{a+{3\over 2}}(\beta^2)
e^{-\beta^2/2},
\end{equation}
where $\Gamma(z)$ stands for the $\Gamma$-function, $n$ is the usual
oscillator quantum number (which should be distinguished from the wobbling
quantum number $n_w$), $L_n^a(z)$ denotes the
Laguerre polynomials \cite{AbrSt}, and
$$ 
a=-{3\over 2} +\sqrt{{4L(L+1)-3\alpha^2+9\over 4}+\beta_0^4 } $$
\begin{equation}\label{eq:e14}
=  -{3\over 2} +\sqrt{{L(L+4)+3n_w(2L-n_w) +9 \over 4} +\beta_0^4 } .
\end{equation}
The energy eigenvalues are then (in $\hbar \omega=1$ units)
$$ 
E_{n,L}^{(n_w)}= 2n+a+{5\over 2} $$
\begin{equation}\label{eq:e15}
= 2n+1 + \sqrt{{L(L+4)+3 n_w (2L-n_w)+9
\over 4} + \beta_0^4 } ,
\end{equation}
where $n=0$,1,2,\dots
One can see that a formal correspondence between the energy levels of the
Z(5) model and the present model, to which we shall refer as the
Z(5)-D model can be
established through the relation $n=s-1$, which expresses just a formal
one-to-one correspondence between the states in the two spectra,
while the origin of the two quantum numbers is different, $s$ labeling the
order of a zero of a Bessel function and $n$ labeling the number of zeros
of a Laguerre polynomial. For the energy states the notation
$E_{s,n_w,L} = E_{n+1,n_w,L}$ will be used, as in Refs. \cite{IacX5,Z5}.
Therefore the ground state band corresponds to $s=1$ ($n=0$) and $n_w=0$.

In the limit $\beta_0\to \infty$ one can expand the square root in Eq.
(\ref{eq:e15}) and keep only the lowest order term, thus obtaining
\begin{equation}\label{eq:e17}
E_L^{(n_w)} = A [L(L+4)+3 n_w (2L-n_w)],
\end{equation}
where $A$ is a constant, which is the spectrum of the triaxial
rotator obtained in Ref. \cite{MtVNPA}.

In the special case $\beta_0=0$, {\it i.e.}, in the case that a harmonic oscillator
is used, one obtains a
parameter-free (up to overall scale factors) exactly
soluble model to which we shall refer as the Z(5)-$\beta^2$ model,
in analogy to the E(5)-$\beta^{2n}$ \cite{Arias,BonE5} and X(5)-$\beta^{2n}$
\cite{BonX5} models. This model represents a triaxial vibrator.

\section{The $\gamma$-part of the spectrum} 

The $\gamma$-part of the spectrum is obtained from Eq. (\ref{eq:e4}),
as described in Ref. \cite{Z5}, by putting in it a harmonic oscillator
potential having a minimum at $\gamma =\pi/6$, i.e.
\begin{equation}\label{eq:e19}
v(\gamma)= {1\over 2} c \left( \gamma-{\pi \over 6}\right)^2 =
{1\over 2} c \tilde \gamma^2, \qquad \tilde \gamma = \gamma -{\pi \over 6}.
\end{equation}
In the case of $\gamma \approx \pi/6$ a simple harmonic oscillator equation
in the variable $\tilde \gamma$ occurs.
Similar potentials and solutions in the $\gamma$-variable have been
considered in \cite{Bohr,Dav24}.

The total energy in the case of the Z(5)-D model is then
$$ 
E(n,n_w,L,n_{\tilde \gamma},\beta_0)= E_0 + A
\left[ 2n+1 \right.$$
\begin{equation}\label{eq:e25}
\left. +\sqrt{ {L(L+4)+3 n_w(2L-n_w)+9 \over 4} +\beta_0^4 }
\right] + B n_{\tilde \gamma},
\end{equation}
where $n_{\tilde \gamma}$ is the number of oscillator quanta in the
$\tilde \gamma$ degree of freedom, and $E_0$, $A$, $B$ are free
parameters.

It should be noticed that in Eq. (\ref{eq:e4})
there is a latent dependence on $s$, $L$, and $n_w$ ``hidden'' in the
$\langle \beta^2 \rangle$ term.
The approximate separation of the $\beta$ and $\gamma$ variables
is achieved by considering an adiabatic limit, as in the X(5) case
\cite{IacX5,Bijker}.

\section{B(E2) transition rates} 

The quadrupole operator is given by
$$ 
T^{(E2)}_\mu = t \beta \left[ {\cal D}^{(2)}_{\mu,0}(\theta_i)\cos\left(\gamma
-{2\pi\over 3}\right) \right. $$
\begin{equation}\label{eq:e31}
\left. +{1\over \sqrt{2}}
({\cal D}^{(2)}_{\mu,2}(\theta_i)+{\cal D}^{(2)}_{\mu,-2}(\theta_i) )
\sin\left(\gamma -{2\pi\over 3} \right) \right],
\end{equation}
where $t$ is a scale factor,
while in the Wigner functions, ${\cal D}^{(2)}$,  the quantum number $\alpha$ appears next
to $\mu$, and the quantity $\gamma -2\pi/3$ in the trigonometric functions
is obtained from $\gamma-2\pi k/3$ for $k=1$, since in the present case
the projection $\alpha$ along the body-fixed $\hat x'$-axis is used.

The symmetrized wave function for Z(5)-D reads
$$ 
\Psi(\beta,\gamma,\theta_i) = \xi_{n,\alpha,L}(\beta) \eta_{n_{\tilde \gamma}}
(\tilde \gamma) $$
\begin{equation}\label{eq:35}
\sqrt{ 2L+1\over 16\pi^2 (1+\delta_{\alpha,0})} ({\cal D}^{(L)}_{\mu,\alpha}
+(-1)^L {\cal D}^{(L)}_{\mu,-\alpha}) ,
\end{equation}
where the normalization factor occurs from the standard integrals
involving two Wigner functions \cite{Edmonds} and is the same as in
\cite{MtVNPA}. $\alpha$ has to be an even integer \cite{MtVNPA},
while for $\alpha=0$ it is clear that only even values of $L$ are
allowed, since the symmetrized wave function is vanishing otherwise.

The calculation of B(E2)s proceeds as in Ref. \cite{Z5} and need not be
repeated here. In the calculation of matrix elements the integral
over $\tilde \gamma$ leads to unity  [because of the normalization
of $\eta(\tilde \gamma)$, and taking into account that $\gamma$ in Eq. (\ref{eq:e31}) 
is fixed to the $\pi/6$ value, because of the steep potential used in $\gamma$],
while the integral over $\beta$ takes the form
$$ 
I_\beta(n_i,L_i,\alpha_i,n_f,L_f,\alpha_f)= $$
\begin{equation}\label{eq:e36}
\int \beta
\xi_{n_i,\alpha_i,L_i}(\beta) \xi_{n_f,\alpha_f,L_f}(\beta) \beta^4 d\beta,
\end{equation}
where the $\beta$ factor comes from Eq. (\ref{eq:e31}), and the $\beta^4$
factor comes from the volume element \cite{Bohr}.
It is worth reminding, though, that a $\Delta \alpha =\pm 2$
selection rule occurs, which results in vanishing quadrupole moments.

\section{Numerical results} 

\subsection{Spectra} 

The lowest bands for the Z(5)-D model are shown in Table~1 for the limiting parameter values
$\beta_0=0$ (the Z(5)-$\beta^2$ model) and $\beta\to\infty$ (the triaxial rotor model \cite{Dav24,MtVNPA}),
as well as for the intermediate value $\beta_0=2$ (for illustative purposes). The levels of Z(5) \cite{Z5}
are also shown for comparison. The bands shown are

i) The ground state band (gsb), with $(s=1, n_w=0)$.

ii) The quasi-$\gamma_1$ band, composed by the even $L$ levels with $(s=1, n_w=2)$
and the odd $L$ levels with $(s=1, n_w=1)$.

iii) The quasi-$\gamma_2$ band, composed by the even $L$ levels with $(s=1, n_w=4)$
and the odd $L$ levels with $(s=1, n_w=3)$.

iv) The quasi-$\beta_1$ band, with $(s=2, n_w=0)$.

v) The quasi-$\beta_2$ band, with $(s=3, n_w=0)$.

Since the last two bands go to infinity for $\beta_0\to \infty$,
the energy levels for $\beta_0=3$ have been shown instead.

In all cases $B=0$ has been used in Eq. (\ref{eq:e25}), {\it i.e.},
the term involving $n_{\bar \gamma}$ has been ignored.

For all bands a uniform raising of the energies from the triaxial vibrator
($\beta_0=0$) values to the triaxial rigid rotator ($\beta_0\to\infty$) values is observed.

A quantity being very sensitive to structural changes (since it is
a discrete derivative of energies) is the
odd--even staggering in gamma bands, described by
quantity \cite{stag}
\begin{equation}\label{eq:stagg}
S(J)={ E(J_\gamma^+)+E((J-2)_\gamma^+)-2 E((J-1)_\gamma^+) \over
E(2_1^+)  },
\end{equation}
which measures the displacement of the $(J-1)_\gamma^+$ level
relative to the average of its neighbors, $J_\gamma^+$ and
$(J-2)_\gamma^+$, normalized to the energy of the first excited
state of the ground state band, $2_1^+$.

It is known \cite{stag2} that $\gamma$-soft shapes exhibit staggering
with negative values at even $L$ and positive values at odd $L$,
while triaxial $\gamma$-rigid shapes exhibit the opposite behavior,
{\it i.e.}, positive values at even $L$ and negative values at odd $L$.
In Table~1 it is clear that the present models exhibit strong staggering
of the triaxial type, with the even-$L$ levels  growing much faster with $L$
than the odd-$L$ levels.

\subsection{Variational procedure} 

A variational procedure appropriate for locating the behaviour of various
physical quantities at a critical point has been introduced
\cite{varPLB,varPRC} and applied for recovering the E(5) \cite{IacE5} and
X(5) \cite{IacX5} ground state
bands from Davidson potentials in the relevant frameworks.
The method is applicable in cases in which one has a one-parameter potential
spanning the region between two limiting symmetries. The method is based on
the fact that if a shape/phase transition occurs between these two symmetries,
the rate of change of various physical quantities should become maximum at the
critical point \cite{Werner}. The parameter value corresponding to the maximum,
$\beta_{0,m}$, is determined for each value of angular momentum separately.

The variational procedure used here resembles the standard Ritz variational procedure
of quantum mechanics \cite{Greiner}, in which a trial wave function containing a free 
parameter is used, while here a potential containing a free parameter is used, 
a difference being that in the Ritz approach the parameter is determined by minimizing 
the energy, while here the parameter is found be maximizing the rate of change of the 
relevant physical quantity. $L$-dependent potentials, like the ones occuring here,
have been used in nuclear physics in optical model potentials \cite{Fiedelday,Mackintosh,Muether},
as well as in the study of quasimolecular resonances \cite{Scheid}.
The method is also analogous to the variable moment of inertia model (VMI) \cite{VMI}, 
in which the energy is minimized with respect to the moment of inertia (which depends on the 
angular momentum) separately for each value of the angular momentm $L$. 

In the present case, as seen in subsec. 2.1, the Davidson potentials of Eq.
(\ref{eq:e0}) lead to a triaxial vibrator Z(5)-$\beta^2$ for
$\beta_0=0$, while they give the rigid triaxial rotator
\cite{DavFil,DavRos,MtVNPA} for $\beta_0\to\infty$. Applying the variational
procedure to the energy ratios $E(L)/E(2)$ of the ground state band
($s=1$, $n_w=0$)
of the Z(5)-D model, where $\beta_0$ is the free parameter
serving to span the region between the two limiting cases,
we are led to the results shown in Table~1, where for each value of the
angular momentum $L$ the location of the maximum, $\beta_{0,m}$, and the
corresponding energy (normalized to the energy of the first excited state)
are given. It is clear that the band determined through the variational
procedure agrees very well with the ground state band of the Z(5) model.
The agreement remains equally good for the $s=1$, $n_w=1$, 2, 3, 4 bands,
also shown in Table~1,
thus indicating that the Z(5) model is possibly related to a shape/phase
transition from a triaxial vibrator to the rigid triaxial rotator.

\subsection{B(E2) transition rates} 

Both intraband and interband B(E2) transition rates for the same models
are reported in Table~2. In addition, results for the O(6) limit of the
Interacting Boson Model \cite{IA} are shown for comparison, derived
from the expressions given in Ref. \cite{IA}, the final results reading
$$ 
R_{g,g}(L+2\to L) = {B(E2; (L+2)_g \to L_g) \over B(E2; 2_g \to 0_g)} $$ 
\begin{equation}\label{eq:e68}
={5\over 2} {(L+2)\over (L+5)} {(2N-L)(2N+L+8)\over 4N(N+4)},
\end{equation}
$$ 
R_{\gamma_{even},g}(L\to L) = {B(E2; L_\gamma \to L_g) \over B(E2;
2_g \to 0_g)}  $$
\begin{equation}\label{eq:e69}
= {10 (L+1) \over (L+5)(2L-1)} {(2N-L)(2N+L+8)\over 4N(N+4)},
\end{equation}
$$
R_{\gamma_{odd},g}(L\to L+1) = {B(E2; L_\gamma \to (L+1)_g) \over
B(E2; 2_g \to 0_g)} $$ 
\begin{equation}\label{eq:e70}
={5(L-1)(2L+3)\over L(L+6)(2L+1)} {(2N-L-1)(2N+L+9)\over 4N(N+4)},
\end{equation}
$$ R_{\gamma_{even} \to \gamma_{even} }(L+2\to L) =
{B(E2; (L+2)_{\gamma} \to L_\gamma) \over B(E2; 2_g \to 0_g)}$$
\begin{equation}\label{eq:e71}
 ={5L(2L+7) \over 2(L+7)(2L+3)} {(2N-L-2)((2N+L+10)\over 4N(N+4)},
\end{equation}
$$  R_{\gamma_{odd} \to \gamma_{odd}}(L+2\to L)=
{B(E2; (L+2)_{\gamma} \to L_\gamma) \over B(E2; 2_g \to 0_g)}$$
\begin{equation}\label{eq:e72}
 ={5(L-1)(L+3)(L+4)\over 2(L+1)(L+2)(L+8)} {(2N-L-3)(2N+L+11)\over 4N(N+4)},
\end{equation}
$$ R_{\gamma_{odd} \to \gamma_{even} }(L\to L-1)  =
{B(E2; L_{\gamma} \to (L-1)_\gamma) \over B(E2; 2_g \to 0_g)} $$
\begin{equation}\label{eq:e73}
= {30 (L+2)\over (L-1)(L+6)(2L+1)} {(2N-L-1)(2N+L+9)\over 4N(N+4)}.
\end{equation}
In all of the above equations, $N$ stands for the boson number.
Numerical results for $N\to \infty$ are reported in Table~2.
We remark that the O(6) predictions for $N\to \infty$ are very similar
to the ones of the rigid triaxial rotator \cite{DavFil,DavRos}, i.e.
to these of the Z(5)-D model for $\beta_0 \to \infty$.

\section{Comparison to experiment} 

As seen from Table~1, one should look for nuclei having
ground state bands charactrized by $R_{4/2}=E(4)/E(2)$ ratios
between 2.150 and 2.667, while the $\gamma_1$ bandhead (normalized
to the $2_1^+$ state) should be between 1.734 and 2.000,
the $\beta_1$ bandhead (normalized in the same way) being above 2.528~.
The Xe isotopes $^{128-132}$Xe, lying below the N=82 shell closure,
nearly fulfil these conditions. Results of one-parameter ($\beta_0$) rms fits
are shown in Table~3, with $\sigma$ being the quality measure
\begin{equation}\label{eq:e99}
\sigma = \sqrt{ { \sum_{i=1}^n (E_i(exp)-E_i(th))^2 \over
(n-1)E(2_1^+)^2} }.
\end{equation}
The overall agreement is good, with the notable exception of the even-$L$ levels
of the quasi-$\gamma_1$ band, which grow too fast, as already remarked at the end
of subsec. 5.1~. As a result, the theoretical predictions exhibit strong triaxial
odd--even staggering, which is not seen experimentally. Indeed, the Xe isotopes are 
known \cite{stag2} to exhibit staggering of the $\gamma$-soft type, in contrast to the 
strong triaxial $\gamma$-rigid staggering shown here by the theoretical values. 
The only nuclei found in the extended recent search of Ref. \cite{stag2}
to possess $\gamma_1$ bands with triaxial shapes  are $^{112}$Ru, $^{170}$Er, $^{192}$Os, $^{192}$Pt, and $^{232}$Th, 
all of them located in the nuclear chart far from the Xe isotopes considered here.  

In Table~4 the existing B(E2) transition rates of the same nuclei are compared to the Z(5)-D
model predictions for the parameter values obtained from fitting the spectra.
No fitting of the B(E2) values has been performed. The theoretical predictions are
in general higher than the experimental values, but in most cases lie within
the experimental error bars, or quite near them.

\section{Conclusions} 

Z(5) \cite{Z5} is a solution of the Bohr Hamiltonian similar to the X(5) \cite{IacX5} solution,
with the notable difference that it regards triaxial shapes ($\gamma \approx \pi/6$)
instead of prolate deformed shapes ($\gamma \approx 0$). Predictions for spectra
and B(E2) transition rates are parameter independent (up to overall scale factors).

In the present Z(5)-D solution, the infinite square well $u(\beta)$ potential, used in Z(5),
is replaced by the Davidson potential \cite{Dav}, involving a free parameter, $\beta_0$.
As a result, Z(5)-D can cover the region between a triaxial vibrator and the rigid triaxial rotator \cite{DavFil,DavRos}.
In addition to providing easily comparable to experiment analytical solutions for spectra and B(E2) values
within this wide region, the present solution has an interesting by-product. Using a variational procedure \cite{varPLB,varPRC}
it is pointed out that the Z(5) solution corresponds to the critical point of the shape phase transition
from a triaxial vibrator to the rigid triaxial rotator. However, the Z(5) solution  is {\sl not} a special case
of Z(5)-D, obtained for a specific parameter value, or a limiting case of Z(5)-D. Using the Davidson potential 
one can cover the whole way from triaxial vibrator to triaxial rotator, but one cannot get the critical point 
as a special case. This is due to the shape of the Davidson potential, which is not flat, as the potential 
is expected to be at the critical point. The same situation has occured in the ESD model \cite{ESD}, in which 
the Davidson potential is used in order to interpolate between a vibrator and the prolate axial rotator 
with $\gamma\approx 0$. Using the ESD model one can obtain very good fits of many nuclei from the prolate 
rotator limit down to close to the critical point, but one cannot describe the nuclei very close to the critical point
\cite{ESD}.  

Concerning the separation of variables which allowed for analytical solutions,
a potential of the form $u(\beta)+v(\gamma)$ has been used, bringing in the
approximations used in X(5) \cite{IacX5}. These approximations can be avoided in two ways:

\noindent i) Using potentials of the form $u(\beta)+v(\gamma)/\beta^2$, which are known \cite{Fort}
to allow for exact separation of variables without any approximations.

\noindent ii) Using the powerful techniques of the Algebraic Collective Model \cite{Rowe735,Rowe753,Rowe79}, which allow
for the exact numerical diagonalization of any Bohr Hamiltonian.

The first path has been used for a detailed study of the Davidson potential plugged in the
Bohr Hamiltonian for $\gamma\approx 0$ \cite{ESD}. The main advantage of this solution
is that all bands are treated on equal footing with respect to the influence of the $v(\gamma)$ potential,
while in the present solution only the quasi-$\gamma$ bands are affected. A similar study
for $\gamma \approx \pi/6$ case would be interesting. The analytical solution and a brief comparison to experiment
in the Os region has already been given in Ref. \cite{FDBH}.

The second path has been recently used for the description of a triaxial symmetry top \cite{Rowe79}, as well as
for the study the onset of rigid triaxial deformation \cite{Caprio672}. Further investigations of triaxial shapes
using this powerful tool should also be revealing.


\begin{table*}

\caption{Energy spectra of the Z(5)-D model (for different values of the parameter
$\beta_0$), and for the Z(5) model \cite{Z5}. $\beta_0=0$ corresponds to the Z(5)-$\beta^2$
model (a triaxial vibrator), while $\beta \to \infty$ is the rigid triaxial
rotator \cite{DavFil,DavRos}. The notation $L_{s,n_w}$ is used.
All levels are measured from the ground state, $0_{1,0}$, and are normalized
to the first excited state, $2_{1,0}$. See subsec. 5.1 for further discussion.
In addition, the energy levels resulting from the variational procedure of subsec. 5.2
are reported (labelled by ``var''), along with the parameter values $\beta_{0,m}$
at which they are obtained. See subsec. 5.2 for further discussion.
}

\bigskip

\begin{tabular}{ r r r r r r r |  r r r r r r r  }
\hline
$\beta_0$ & 0 & 2 & $\infty$ &               &     &   & $\beta_0$ & 0 & 2 & $\infty$ &               &     &     \\
 $L_{s,n_w}$ &   &    &      & $\beta_{0,m}$ & var & Z(5) & $L_{s,n_w}$ &   &    &      & $\beta_{0,m}$ & var & Z(5)\\
\hline
 $0_{1,0}$ & 0.000 & 0.000 & 0.000 &      &       & 0.000 &  &       &       &       &      &       & \\
 $2_{1,0}$ & 1.000 & 1.000 & 1.000 &      &       & 1.000 &  &       &       &       &      &       & \\
 $4_{1,0}$ & 2.150 & 2.521 & 2.667 &1.375 & 2.341 & 2.350 &  &       &       &       &      &       & \\
 $6_{1,0}$ & 3.353 & 4.424 & 5.000 &1.474 & 3.956 & 3.984 &  &       &       &       &      &       & \\
 $8_{1,0}$ & 4.579 & 6.596 & 8.000 &1.562 & 5.819 & 5.877 &  &       &       &       &      &       & \\
$10_{1,0}$ & 5.817 & 8.957 &11.667 &1.640 & 7.915 & 8.019 &  &       &       &       &      &       & \\
$12_{1,0}$ & 7.063 &11.450 &16.000 &1.713 &10.237 &10.403 &  &       &       &       &      &       & \\
$14_{1,0}$ & 8.313 &14.039 &21.000 &1.780 &12.781 &13.024 &  &       &       &       &      &       & \\
$16_{1,0}$ & 9.566 &16.698 &26.667 &1.843 &15.544 &15.878 &  &       &       &       &      &       & \\
$18_{1,0}$ &10.821 &19.410 &33.000 &1.902 &18.523 &18.964 &  &       &       &       &      &       & \\
$20_{1,0}$ &12.077 &22.163 &40.000 &1.960 &21.719 &22.279 &  &       &       &       &      &       & \\
\hline
           &       &       &       &      &       &        & $2_{1,2}$ & 1.734 & 1.932 & 2.000 & 1.336 & 1.833 &  1.837 \\
 $3_{1,1}$ & 2.343 & 2.807 & 3.000 & 1.392 & 2.586 & 2.597 & $4_{1,2}$ & 3.649 & 4.930 & 5.667 & 1.496 & 4.386 &  4.420 \\
 $5_{1,1}$ & 3.791 & 5.177 & 6.000 & 1.507 & 4.597 & 4.634 & $6_{1,2}$ & 5.281 & 7.917 &10.000 & 1.607 & 6.981 &  7.063 \\
 $7_{1,1}$ & 5.169 & 7.703 & 9.667 & 1.600 & 6.790 & 6.869 & $8_{1,2}$ & 6.791 &10.898 &15.000 & 1.697 & 9.713 &  9.864 \\
 $9_{1,1}$ & 6.511 &10.333 &14.000 & 1.681 & 9.182 & 9.318 & $10_{1,2}$ & 8.234 &13.874 &20.667 & 1.776 &12.615 &12.852 \\
$11_{1,1}$ & 7.832 &13.035 &19.000 & 1.754 &11.778 &11.989 & $12_{1,2}$ & 9.635 &16.847 &27.000 & 1.846 &15.703 &16.043 \\
$13_{1,1}$ & 9.140 &15.788 &24.667 & 1.822 &14.581 &14.882 & $14_{1,2}$ &11.008 &19.818 &34.000 & 1.911 &18.986 &19.443 \\
$15_{1,1}$ &10.438 &18.579 &31.000 & 1.885 &17.593 &18.000 & $16_{1,2}$ &12.360 &22.787 &41.667 & 1.972 &22.468 &23.056 \\
$17_{1,1}$ &11.730 &21.399 &38.000 & 1.944 &20.815 &21.341 & $18_{1,2}$ &13.698 &25.755 &50.000 & 2.030 &26.154 &26.884 \\
$19_{1,1}$ &13.017 &24.241 &45.667 & 2.001 &24.248 &24.905 & $20_{1,2}$ &15.024 &28.721 &59.000 & 2.085 &30.046 &30.928 \\
\hline
           &       &       &       &       &       &       & $4_{1,4}$ & 4.066 & 5.663 & 6.667 & 1.526 & 5.012 & 5.056 \\
 $5_{1,3}$ & 4.939 & 7.268 & 9.000 & 1.585 & 6.406 & 6.476 & $6_{1,4}$ & 6.221 & 9.753 &13.000 & 1.665 & 8.644 & 8.767 \\
 $7_{1,3}$ & 6.699 &10.711 &14.667 & 1.692 & 9.537 & 9.683 & $8_{1,4}$ & 8.075 &13.541 &20.000 & 1.767 &12.282 &12.508 \\
 $9_{1,3}$ & 8.313 &14.039 &21.000 & 1.780 &12.781 &13.024 & $10_{1,4}$ & 9.773 &17.143 &27.667 & 1.853 &16.021 &16.372\\
$11_{1,3}$ & 9.841 &17.289 &28.000 & 1.856 &16.180 &16.536 & $12_{1,4}$ &11.374 &20.618 &36.000 & 1.928 &19.904 &20.396\\
$13_{1,3}$ &11.314 &20.486 &35.667 & 1.925 &19.752 &20.237 & $14_{1,4}$ &12.910 &24.003 &45.000 & 1.996 &23.953 &24.598\\
$15_{1,3}$ &12.747 &23.642 &44.000 & 1.989 &23.509 &24.137 & $16_{1,4}$ &14.399 &27.320 &54.667 & 2.059 &28.182 &28.991\\
$17_{1,3}$ &14.152 &26.768 &53.000 & 2.049 &27.460 &28.241 & $18_{1,4}$ &15.853 &30.586 &65.000 & 2.118 &32.601 &33.581\\
$19_{1,3}$ &15.536 &29.871 &62.667 & 2.105 &31.611 &32.553 & $20_{1,4}$ &17.281 &33.811 &76.000 & 2.174 &37.217 &38.373\\
\hline
$\beta_0$ & 0 & 2 & 3 &               &     &  &  $\beta_0$ & 0 & 2 & 3 &               &     &     \\
 $L_{s,n_w}$ &   &  &        & $\beta_{0,m}$ & var & Z(5) &  $L_{s,n_w}$ &   &  &        & $\beta_{0,m}$ & var & Z(5) \\
\hline
 $0_{2,0}$ & 2.528 & 5.921&12.274&  &  & 3.913 & $0_{3,0}$ & 5.055 &11.842&24.548&  &  & 9.782 \\
 $2_{2,0}$ & 3.528 & 6.921&13.274&  &  & 5.697 & $2_{3,0}$ & 6.055 &12.842&25.548&  &  &12.343 \\
 $4_{2,0}$ & 4.678 & 8.442&14.903&  &  & 7.962 & $4_{3,0}$ & 7.205 &14.363&27.177&  &  &15.506 \\
 $6_{2,0}$ & 5.881 &10.345&17.110&  &  &10.567 & $6_{3,0}$ & 8.408 &16.266&29.384&  &  &19.059 \\
 $8_{2,0}$ & 7.107 &12.517&19.835&  &  &13.469 & $8_{3,0}$ & 9.634 &18.439&32.109&  &  &22.933 \\
$10_{2,0}$ & 8.345 &14.878&23.015&  &  &16.646 & $10_{3,0}$ &10.873 &20.799&35.289&  &  &27.103\\
$12_{2,0}$ & 9.590 &17.371&26.588&  &  &20.088 & $12_{3,0}$ &12.118 &23.292&38.862&  &  &31.552\\
$14_{2,0}$ &10.840 &19.960&30.497&  &  &23.788 & $14_{3,0}$ &13.368 &25.881&42.771&  &  &36.272\\
$16_{2,0}$ &12.093 &22.619&34.692&  &  &27.740 & $16_{3,0}$ &14.621 &28.540&46.966&  &  &41.258\\
$18_{2,0}$ &13.348 &25.331&39.129&  &  &31.942 & $18_{3,0}$ &15.876 &31.253&51.403&  &  &46.504\\
$20_{2,0}$ &14.605 &28.084&43.772&  &  &36.390 & $20_{3,0}$ &17.132 &34.005&56.046&  &  &52.007\\
\hline
\end{tabular}
\end{table*}


\begin{table*}

\caption{Intraband and interband B(E2) transition rates, normalized to the one
between the two lowest states, B(E2;$2_{1,0}\to 0_{1,0}$), are given for Z(5)-D model
(for different values of the parameter $\beta_0$), for the Z(5) model \cite{Z5},
and for the O(6) limit of the Interacting Boson Model \cite{IA}.
$\beta_0=0$ corresponds to the Z(5)-$\beta^2$ model (a triaxial vibrator),
while $\beta \to \infty$ is the rigid triaxial rotator \cite{DavFil,DavRos}.
The notation $L_{s,n_w}$ is used, while the initial state is labelled by $(i)$,
and the final state by $(f)$.
}

\bigskip

\begin{tabular}{ r r r r r r r | r r r r r r r}
\hline
$\beta_0$ &   &   0  & 2 & $ \infty$ &  &  & $\beta_0$ &   &   0  & 2 & $ \infty$ &  & \\
$L^{(i)}_{s,n_w}$ & $L^{(f)}_{s,n_w}$ &   &     &    & O(6) & Z(5) & $L^{(i)}_{s,n_w}$ & $L^{(f)}_{s,n_w}$ &   &     &    & O(6) & Z(5)\\
\hline
 $2_{1,0}$ & $0_{1,0}$ & 1.000 & 1.000 & 1.000 & 1.000 & 1.000 & $2_{2,0}$ & $0_{2,0}$ & 1.480 & 1.284 & 1.000 &       & 0.774 \\
 $4_{1,0}$ & $2_{1,0}$ & 1.834 & 1.493 & 1.389 & 1.429 & 1.590 & $4_{2,0}$ & $2_{2,0}$ & 2.440 & 1.865 & 1.389 &       & 1.192 \\
 $6_{1,0}$ & $4_{1,0}$ & 2.919 & 2.041 & 1.731 & 1.667 & 2.203 & $6_{2,0}$ & $4_{2,0}$ & 3.647 & 2.477 & 1.731 &       & 1.643 \\
 $8_{1,0}$ & $6_{1,0}$ & 3.955 & 2.497 & 1.912 & 1.818 & 2.635 & $8_{2,0}$ & $6_{2,0}$ & 4.746 & 2.955 & 1.912 &       & 1.975 \\
$10_{1,0}$ & $8_{1,0}$ & 4.976 & 2.934 & 2.024 & 1.923 & 2.967 & $10_{2,0}$ & $8_{2,0}$ & 5.804 & 3.398 & 2.024 &       & 2.242 \\
$12_{1,0}$ &$10_{1,0}$ & 5.989 & 3.370 & 2.100 & 2.000 & 3.234 & $12_{2,0}$ &$10_{2,0}$ & 6.844 & 3.836 & 2.100 &       & 2.466 \\
$14_{1,0}$ &$12_{1,0}$ & 6.999 & 3.810 & 2.155 & 2.059 & 3.455 & $14_{2,0}$ &$12_{2,0}$ & 7.874 & 4.277 & 2.155 &       & 2.660 \\
$16_{1,0}$ &$14_{1,0}$ & 8.007 & 4.257 & 2.197 & 2.105 & 3.642 & $16_{2,0}$ &$14_{2,0}$ & 8.897 & 4.723 & 2.197 &       & 2.829 \\
$18_{1,0}$ &$16_{1,0}$ & 9.014 & 4.708 & 2.230 & 2.143 & 3.803 & $18_{2,0}$ &$16_{2,0}$ & 9.915 & 5.175 & 2.230 &       & 2.980 \\
$20_{1,0}$ &$18_{1,0}$ &10.020 & 5.164 & 2.256 & 2.174 & 3.944 & $20_{2,0}$ &$18_{2,0}$ &10.931 & 5.630 & 2.256 &       & 3.115 \\
\hline
 $4_{1,2}$ & $2_{1,2}$ & 0.912 & 0.682 & 0.595 & 0.873 & 0.736 & $3_{1,1}$ & $2_{1,2}$ & 2.731 & 2.006 & 1.786 & 1.190 & 2.171 \\
 $6_{1,2}$ & $4_{1,2}$ & 1.583 & 0.986 & 0.734 & 1.240 & 1.031 & $5_{1,1}$ & $4_{1,2}$ & 1.991 & 1.244 & 0.955 & 0.434 & 1.313 \\
 $8_{1,2}$ & $6_{1,2}$ & 2.816 & 1.617 & 1.051 & 1.462 & 1.590 & $7_{1,1}$ & $6_{1,2}$ & 2.183 & 1.262 & 0.851 & 0.231 & 1.260 \\
$10_{1,2}$ & $8_{1,2}$ & 4.038 & 2.215 & 1.278 & 1.614 & 2.035 & $9_{1,1}$ & $8_{1,2}$ & 2.247 & 1.241 & 0.746 & 0.145 & 1.164 \\
$12_{1,2}$ &$10_{1,2}$ & 5.231 & 2.786 & 1.446 & 1.726 & 2.394 & $11_{1,1}$ &$10_{1,2}$ & 2.265 & 1.214 & 0.657 & 0.100 & 1.069 \\
$14_{1,2}$ &$12_{1,2}$ & 6.395 & 3.339 & 1.574 & 1.813 & 2.690 & $13_{1,1}$ &$12_{1,2}$ & 2.265 & 1.189 & 0.585 & 0.073 & 0.984 \\
$16_{1,2}$ &$14_{1,2}$ & 7.535 & 3.877 & 1.674 & 1.882 & 2.938 & $15_{1,1}$ &$14_{1,2}$ & 2.258 & 1.168 & 0.526 & 0.056 & 0.910 \\
$18_{1,2}$ &$16_{1,2}$ & 8.655 & 4.406 & 1.755 & 1.938 & 3.151 & $17_{1,1}$ &$16_{1,2}$ & 2.247 & 1.149 & 0.478 & 0.044 & 0.846 \\
$20_{1,2}$ &$18_{1,2}$ & 9.759 & 4.927 & 1.822 & 1.985 & 3.335 & $19_{1,1}$ &$18_{1,2}$ & 2.236 & 1.133 & 0.437 & 0.036 & 0.790 \\
\hline
 $2_{1,2}$ & $2_{1,0}$ & 1.865 & 1.520 & 1.429 & 1.429 & 1.620 & $3_{1,1}$ & $4_{1,0}$ & 1.618 & 1.147 & 1.000 & 0.476 & 1.243 \\
 $4_{1,2}$ & $4_{1,0}$ & 0.459 & 0.323 & 0.273 & 0.794 & 0.348 & $5_{1,1}$ & $6_{1,0}$ & 1.449 & 0.917 & 0.714 & 0.430 & 0.972 \\
 $6_{1,2}$ & $6_{1,0}$ & 0.292 & 0.187 & 0.143 & 0.579 & 0.198 & $7_{1,1}$ & $8_{1,0}$ & 1.351 & 0.796 & 0.556 & 0.374 & 0.808 \\
 $8_{1,2}$ & $8_{1,0}$ & 0.211 & 0.127 & 0.088 & 0.462 & 0.129 & $9_{1,1}$ &$10_{1,0}$ & 1.287 & 0.724 & 0.455 & 0.327 & 0.696 \\
$10_{1,2}$ &$10_{1,0}$ & 0.165 & 0.094 & 0.059 & 0.386 & 0.092 & $11_{1,1}$ &$12_{1,0}$ & 1.243 & 0.676 & 0.385 & 0.291 & 0.614 \\
$12_{1,2}$ &$12_{1,0}$ & 0.135 & 0.074 & 0.043 & 0.332 & 0.069 & $13_{1,1}$ &$14_{1,0}$ & 1.211 & 0.643 & 0.333 & 0.261 & 0.551 \\
$14_{1,2}$ &$14_{1,0}$ & 0.114 & 0.061 & 0.032 & 0.292 & 0.054 & $15_{1,1}$ &$16_{1,0}$ & 1.186 & 0.619 & 0.294 & 0.237 & 0.507 \\
$16_{1,2}$ &$16_{1,0}$ & 0.099 & 0.052 & 0.025 & 0.261 & 0.043 & $17_{1,1}$ &$18_{1,0}$ & 1.167 & 0.601 & 0.263 & 0.216 & 0.459 \\
$18_{1,2}$ &$18_{1,0}$ & 0.087 & 0.045 & 0.020 & 0.236 & 0.035 & $19_{1,1}$ &$20_{1,0}$ & 1.151 & 0.587 & 0.238 & 0.199 & 0.425 \\
$20_{1,2}$ &$20_{1,0}$ & 0.078 & 0.040 & 0.017 & 0.215 & 0.030 &            &           &       &       &       &       &       \\
\hline
 $5_{1,1}$ & $3_{1,1}$ & 1.667 & 1.147 & 0.955 & 0.955 & 1.235 &            &           &       &       &       &       &       \\
 $7_{1,1}$ & $5_{1,1}$ & 2.891 & 1.778 & 1.310 & 1.319 & 1.851 &            &           &       &       &       &       &       \\
 $9_{1,1}$ & $7_{1,1}$ & 4.061 & 2.338 & 1.535 & 1.528 & 2.308 &            &           &       &       &       &       &        \\
$11_{1,1}$ & $9_{1,1}$ & 5.191 & 2.865 & 1.690 & 1.668 & 2.665 &            &           &       &       &       &       &        \\
$13_{1,1}$ &$11_{1,1}$ & 6.292 & 3.374 & 1.802 & 1.771 & 2.952 &            &           &       &       &       &       &        \\
$15_{1,1}$ &$13_{1,1}$ & 7.373 & 3.873 & 1.887 & 1.850 & 3.190 &            &           &       &       &       &       &        \\
$17_{1,1}$ &$15_{1,1}$ & 8.440 & 4.366 & 1.954 & 1.913 & 3.392 &            &           &       &       &       &       &        \\
$19_{1,1}$ &$17_{1,1}$ & 9.496 & 4.856 & 2.007 & 1.965 & 3.566 &            &           &       &       &       &       &        \\
\hline

\end{tabular}
\end{table*}


\begin{table}

\caption{Comparison of theoretical predictions of the Z(5)-D model, labelled
by the relevant $\beta_0$ value, to experimental spectra
of $^{128}$Xe \cite{Xe128}, $^{130}$Xe \cite{Xe130}, and $^{132}$Xe \cite{Xe132}.
In each column all energies are normalized to the energy of the relevant $2_1^+$ state.
The quality measure $\sigma$ of Eq. (\ref{eq:e99}) is used.
See Sec. 6 for further discussion.}

\bigskip

\begin{tabular}{ r l l l l l l }
\hline
             & $^{128}$Xe & $^{128}$Xe & $^{130}$Xe & $^{130}$Xe & $^{132}$Xe & $^{132}$Xe \\
 $L_{s,n_w}$ & exp        & $\beta_0=1.32$  & exp   &  $\beta_0=1.11$    & exp & $\beta_0=0$ \\
\hline
 $4_{1,0}$ & 2.333 & 2.323 & 2.247 & 2.255 & 2.157 & 2.150 \\
 $6_{1,0}$ & 3.922 & 3.805 & 3.627 & 3.621 & 3.163 & 3.353 \\
 $8_{1,0}$ & 5.674 & 5.372 & 5.031 & 5.040 &       &       \\
$10_{1,0}$ & 7.597 & 6.986 & 6.457 & 6.489 &       &       \\
$12_{1,0}$ &       &       & 7.867 & 7.956 &       &       \\
$14_{1,0}$ &       &       & 9.458 & 9.434 &       &       \\
           &       &       &       &       &       &       \\
 $2_{1,2}$ & 2.189 & 1.830 & 2.093 & 1.793 & 1.944 & 1.734 \\
 $4_{1,2}$ & 3.620 & 4.180 & 3.373 & 3.961 & 2.940 & 3.649 \\
 $6_{1,2}$ & 5.150 & 6.284 &       &       &       &       \\
           &       &       &       &       &       &       \\
 $3_{1,1}$ & 3.228 & 2.555 & 3.045 & 2.471 & 2.701 & 2.343 \\
 $5_{1,1}$ & 4.508 & 4.360 & 4.051 & 4.125 & 3.246 & 3.791 \\
 $7_{1,1}$ & 6.165 & 6.138 &       &       &       &       \\
           &       &       &       &       &       &       \\
 $0_{2,0}$ & 3.574 & 3.452 & 3.346 & 3.028 & 2.771 & 2.528 \\
 $2_{2,0}$ & 4.515 & 4.452 &       &       &       &       \\
\hline
$\sigma$   &       & 0.495 &       & 0.297 &       & 0.422 \\
\hline
\end{tabular}
\end{table}


\begin{table}

\caption{Comparison of theoretical predictions of the Z(5)-D model, labelled
by the relevant $\beta_0$ value, to experimental B(E2) values
of $^{128}$Xe \cite{Xe128} and $^{132}$Xe \cite{Xe132}.
In each column all B(E2)s  are normalized to the  $2_1^+\to 0_1^+$ transition.
See Sec. 6 for further discussion.
}

\bigskip

\begin{tabular}{ r r l l l l  }
\hline
       &       & $^{128}$Xe & $^{128}$Xe  & $^{132}$Xe & $^{132}$Xe \\
 $L_{s,n_w}^{(i)}$ & $L_{s,n_w}^{(f)}$ & exp & $\beta_0=1.32$ & exp & $\beta_0=0$ \\
\hline
 $4_{1,0}$ & $2_{1,0}$ & $1.468\pm 0.201$ & 1.648 & $1.238\pm 0.180$ & 1.834 \\
 $6_{1,0}$ & $4_{1,0}$ & $1.940\pm 0.275$ & 2.464 &                  &       \\
 $8_{1,0}$ & $6_{1,0}$ & $2.388\pm 0.398$ & 3.228 &                  &       \\
           &           &                  &       &                  &       \\
 $2_{1,2}$ & $2_{1,0}$ & $1.194\pm 0.187$ & 1.673 & $1.775\pm 0.288$ & 1.865 \\
 $2_{1,2}$ & $0_{1,0}$ & $0.016\pm 0.002$ & 0.000 & $0.003\pm 0.001$ & 0.000 \\
\hline
\end{tabular}
\end{table}

\end{document}